\documentclass[aps,prb,twocolumn,preprintnumbers,amsmath,amssymb]{revtex4}

\usepackage{graphicx}
\usepackage{amsfonts}
\usepackage{dcolumn}
\usepackage{bm}
\usepackage{float}

\newcommand{\rmH}{\ensuremath{\mathrm{H}}}
\newcommand{\rmT}{\ensuremath{\mathrm{T}}}

\newcommand{\sbr}{\bar{\sigma}}
\newcommand{\dr}{\mathrm{d}r}

\begin{document}

\title{Modulated phases in magnetic models frustrated by long-range interactions}

\author{Erik Nielsen$^1$}
\author{R. N. Bhatt$^{1,2}$}
\author{David A. Huse$^{3}$}
\affiliation{$^1$Department of Electrical Engineering, Princeton University, Princeton, NJ 08544-5263} 
\affiliation{$^2$Princeton Center for Theoretical Physics, Jadwin Hall, Princeton, NJ 08544}
\affiliation{$^3$Department of Physics, Princeton University,
Princeton, NJ 08544}

\date{October 5, 2007}

\begin{abstract}
We study an Ising model in one dimension with short range
ferromagnetic and long range (power law) antiferromagnetic
interactions.  We show that the zero temperature phase diagram in a (longitudinal)
field $\rmH$ involves a sequence of up and down domains whose size varies
continuously with $\rmH$, between $-\rmH_c$ and $\rmH_c$ which
represent the edge of the ferromagnetic up and down phases. The
implications of long range interaction in many body systems are discussed.
\end{abstract}


\maketitle

\section{Introduction}
It is known that long range interactions lead to significant changes
in the behavior of interacting many body systems.  Thus, though the Ising
model in one dimension with a short range ferromagnetic interaction does not exhibit a phase transition, the corresponding model with a ferromagnetic interaction that falls off as a power law $1/r^\lambda$ has a phase
transition at non-zero temperature \cite{Chakravarty} for $\lambda \leq 2$.
A similar effect is found for the Ising spin glass in one
dimension, which is unfrustrated with nearest neighbor interactions, but becomes sufficiently frustrated with a power law interaction so that for $\lambda < 1$ a finite $\mathrm{T}$ transition is attained\cite{KotliarSteinAnderson,BhattAndYoung}.

For short range models with a finite transition temperature, addition of long range interactions having a power-law fall off can lead to changes in the universality class of the phase transition.  For sufficiently small power law exponent, critical exponents are found to vary continuously with the power law exponent\cite{MEF}.

There have been several studies of the Ising model in higher dimensions, frustrated by Coulomb\cite{LowEmery,ViotTarjus} or dipolar\cite{MuMa,NgVanderbilt} long range interactions, but without a magnetic field.  Here we examine
the effects of a frustrating long range interaction on the phase
diagram of a 1D Ising model at zero temperature \emph{in the presence of a
magnetic field}. This work is motivated by the proposal by Spivak and
Kivelson \cite{Spivak2004} (and generalized with Jamei\cite{JameiKivelsonSpivak}) that the putative first order phase
transition between the Wigner crystal and Fermi liquid phases of the
interacting electron gas in two dimensions at $\rmT=0$ is
pre-empted, due to the long-range Coulomb interaction, by a series
of ``microemulsion'' phases characterized by phase separation on a
mesoscopic scale.
In general, a system with a long range interaction that frustrates
the order favored by a short range interaction will not
macroscopically separate into the phases of the unfrustrated system,
once the long range interaction is strong enough.  This is because
the short range interaction energy increase due to having mesoscopic
domains is smaller than the long range interaction energy increase
due to having macroscopic domains.  (Thus, in such systems, the
Maxwell construction for determining phase separation must be generalized\cite{Lorenzana2001}.)

In this work, we demonstrate that a Coulomb frustrated Ising model
in a spatial continuum in one dimension (and its generalization to
other power laws) admits analytical solution at zero temperature. We
find that this system possesses a regime exhibiting modulated phases (\emph{i.e.} mesoscopic phase
separation), with a period that varies continuously with applied magnetic field.

\section{The Model}
We study a one-dimensional frustrated Ising model given by
adding to the familiar ferromagnetic Ising chain in a magnetic field,
where only nearest neighbors interact, a competing long-range
antiferromagnetic interaction with a power law fall-off.  We
investigate the model in a one-dimensional continuum, where its
Hamiltonian is given by:
\begin{eqnarray}
\mathcal{H}_H &=& J\int\limits_{-\infty}^{\infty} \dr\,\left|\frac{d\sigma(r)}{dr}\right| - \rmH \int\limits_{-\infty}^{\infty} \dr\,\sigma(r) \nonumber\\
& & + \frac{Q}{2}\int\limits_{-\infty}^{\infty} \dr_i \dr_j\,
v(r)\sigma(r_i) \sigma(r_j)  \,.\label{eq_Ham1}
\end{eqnarray}
Here $\sigma(r)=\pm 1$ is a bi-valued function of the position $r$ (Ising spin),
$J$ and $Q$ are parameters representing the strengths of the
short-range ferromagnetic and long-range antiferromagnetic
interactions respectively, and $\rmH$ is the strength of a uniform
magnetic field.  In this paper, we take the long-range
antiferromagnetic interaction as $v(r)=1/(r+a)^\lambda$, with
exponent $\lambda
> 0$, and an ultraviolet cutoff $a$ that must
sometimes be retained to eliminate divergences. In the case
$\rmH=0$, this model has been solved by Giuliani {\it et al.}
\cite{Giuliani2006}, and for $\rmH=0$ and $v(r)$ equal to the inverse Fourier transform of the inverse Laplacian, Grousson {\it et al.} \cite{Grousson2000}~have analytically solved
the model in three dimensions.

It proves helpful to perform a Legendre transformation on the energy
and work at fixed average spin density $\sbr$ instead of fixed
magnetic field strength $\rmH$, especially for $\lambda\leq 1$ where
the interaction energy density is infrared divergent for $\sbr\neq
0$. Since the field term in $\mathcal{H}_H$ is constant for fixed
$\sbr$, the Hamiltonian at a given fixed $\sbr$ is:
\begin{equation}
\mathcal{H}_{\sbr} = J\int\limits_{-\infty}^{\infty}
\dr\,\left|\frac{d\sigma(r)}{dr}\right| +
\frac{Q}{2}\int\limits_{-\infty}^{\infty}\dr_i \dr_j\,
\frac{\sigma(r_i)\sigma(r_j)}{(|r_i-r_j| + a)^\lambda}~.
\label{eq_Ham2}
\end{equation}

\section{Details of Calculation}
To investigate the $\mathrm{T}=0$ properties of this classical
system, we minimize the energy density to find the ground state. We
\emph{assume} 
that the ground state has a simple periodic structure, where each
period is comprised of a length $l_\uparrow$ of up spins followed by a
length $l_\downarrow$ of down spins.  In the $\rmH=0$ case, it has been
proved \cite{Giuliani2006} that the ground state must be of this
form, with $l_\uparrow=l_\downarrow$.  In appendix \ref{appMonteCarlo1D}, Monte Carlo results
are presented that confirm a simple periodic configuration to be the ground state
for $\rmH \ne 0$.  A period has total length
$L \equiv l_\uparrow+l_\downarrow$, and simple algebra reveals that $l_\uparrow =
(1+\sbr)L/2$ and $l_\downarrow = (1-\sbr)L/2$. Minimizing the energy density
for a given $\sbr$ under this assumption is equivalent to minimizing
the energy density of a single period with respect to variation in
$L$ (we choose $L$, but any of the interdependent variables \{$l_\uparrow$,
$l_\downarrow$, $L$\} could be used).  
The function $\sigma(r)$ is specified by the two parameters, $\sbr$
and $L$:
\begin{equation}
\sigma(r) = \left\{
\begin{array}{cc}
+1 & 0 < (r\mbox{ mod } L) < l_\uparrow  \\
-1 & l_\uparrow < (r\mbox{ mod } L) < L\,,
\end{array} \right.
\end{equation}
with $l_\uparrow$ as given above.  Before writing down an explicit formula
for the energy density we must choose a zero of energy, and the
appropriate choice depends on the value of $\lambda$.

\subsection{Case I. \,$\lambda<1$}

When $\lambda < 1$, we choose the zero of energy to be a uniform
spin density of value $\sbr$.  This is equivalent to placing the
system in a background jellium of ``spin charge'', with density
$-\sbr$, and results in the replacement of $\sigma$ with
\begin{equation}
\sigma'(r)\equiv \sigma(r) - \sbr
\end{equation}
in (\ref{eq_Ham2}).  The energy density of a configuration with
average spin $\sbr$ and total period $L$ is given by:
\begin{equation}
\epsilon(\sbr,L) = \frac{4J}{L} + \lim_{X\rightarrow\infty}
\frac{Q}{4X}\int_{-X}^X\dr\dr' \,
\frac{\sigma'(r)\sigma'(r')}{(|r-r'|+a)^\lambda}\,.
\label{eq_Edensity1}
\end{equation}
\noindent The first term is the energy of two domain walls per
period divided by the period length $L$.  The second term is the
limit as $X$ goes to infinity, of the energy density due to the
long-range interaction of a finite system of length $2X$; our zero
of energy here has been chosen to make this limit finite. The
function $\sigma'$ is periodic with period $L$, and we define its
Fourier transform:
\begin{eqnarray}
\sigma'(r) &=& \sum_G \sigma'_G e^{irG}\,, \\
\sigma'_G &=& \frac{1}{L}\int_0^L \sigma'(r) e^{-irG} \dr\,,
\end{eqnarray}
where the sum is over reciprocal lattice vectors $G = 2\pi m/L$ for
$m \in \mathbb{Z}$.  Taking the Fourier transform of the second term
in (\ref{eq_Edensity1}) gives
\begin{equation}
\epsilon(\sbr,L) = \frac{4J}{L} + \frac{Q}{2}\sum_G v_G
|\sigma'_G|^2\,. \label{eq_Edensity2}
\end{equation}
We have used that $\int \dr\, e^{ir(G_1-G_2)} = 2X\delta_{G_1,G_2}$,
taken the limit $X\rightarrow \infty$, and have defined the Fourier
transform of $v(r)$, $v_G = \int_{-\infty}^\infty \dr \,
v(r)e^{-irG}$. In the case $\lambda < 1$ we can compute $v_G$ when
$v(r)=1/(r+a)^\lambda$ to obtain:
\begin{eqnarray}
v_G & = & 2\int_0^\infty \frac{\cos(Gr)}{(r+a)^\lambda}\dr \nonumber\\
     & = & 2G^{\lambda-1} \Gamma(1-\lambda) \sin\left(\frac{\pi\lambda}{2}\right) \qquad (\lambda < 1)~, \label{eq_vg}
\end{eqnarray}
where we have skipped intermediate steps in the integration, and in
the second line have let $a \rightarrow 0$.  Looking back to
(\ref{eq_Edensity2}), we next must calculate the Fourier series
coefficients of $\sigma'(r)$. We achieve this by calculating the
Fourier coefficients $\sigma_G$ of $\sigma(r)$.  Then, since
$\sigma'_G = \sigma_G$ for $G \ne 0$, and $\sigma'_{G=0} = 0$ by
definition, we obtain $\sigma'_G$.
\begin{eqnarray}
\sigma_G &=& \frac{1}{L} \int_{-l_\uparrow}^0 \dr \, (1) e^{-irG} + \int_0^{l_\downarrow}\dr (-1)e^{-irG} \nonumber\\
&=& \frac{i}{GL}\left(2-e^{il_\uparrow G} - e^{-il_\downarrow G} \right)
\label{eq_sigmaG}
\end{eqnarray}
By substituting $G=2\pi m/L$, and inserting (\ref{eq_vg}) and
(\ref{eq_sigmaG}) into (\ref{eq_Edensity2}), we arrive at a final
expression for the energy density:
\begin{equation}
\epsilon(\sbr,L) = \frac{4J}{L} + \frac{Q\,2^{\lambda+1}\Gamma(3-\lambda) \sin\frac{\pi \lambda}{2}}{
L^{\lambda-1}(1-\lambda)(2-\lambda)} C(\sbr,\lambda)\,,
\label{eq_Edensity3}
\end{equation}
where we have 
defined
\begin{equation}
C(\sbr,\lambda) = \sum_{m=1}^\infty  \frac{1 - (-1)^m\cos(\pi m\sbr)
}{(\pi m)^{3-\lambda}}~, \label{eq_C}
\end{equation}
which converges for $\lambda < 2$.  Finally, we solve
$\frac{\partial\epsilon}{\partial L}=0$ to obtain the length of the
period ($L$) that minimizes the energy density subject to the
specified value of $\sbr$ (and $\lambda < 1$) in the limit $a
\rightarrow 0$:
\begin{equation}
L_0 = \left( (2-\lambda)\frac{J}{Q}\left[ 2^{\lambda-1}
\Gamma(3-\lambda) \sin\frac{\pi \lambda}{2} C(\sbr,\lambda)
\right]^{-1} \right)^{\frac{1}{2-\lambda}}\,.
\label{eq_optimalCase1}
\end{equation}

\subsection{Case II. \,$1<\lambda <2$}
When $\lambda > 1$ it is best to choose the fully polarized
(ferromagnetic) state as the zero of energy in order to eliminate
the {\it ultraviolet} divergence of the energy integrals.  The
energy density for $\lambda > 1$ then reads:
\begin{eqnarray}
\epsilon(\sbr,L) &=& \frac{4J}{L} + \lim_{X\rightarrow\infty}\frac{Q}{4X}\int\limits_{-X}^X\dr\dr' \, v\left(|r-r'|\right)\nonumber\\
& &  \times \left( \sigma(r)\sigma(r')-1 \right)\,.
\label{eq_Edensity10}
\end{eqnarray}
The first term is unchanged from the previous case, since the
derivative is only sensitive to domain boundaries, and in the second
term we explicitly subtract the energy of a fully polarized
configuration.  Performing a Fourier transform on the second term
leads to:
\begin{equation}
\epsilon(\sbr,L) = \frac{4J}{L} + \frac{Q}{2} \sum_{G\ne0}
\left(v_G-v_0\right) |\sigma_G|^2\,. \label{eq_Edensity11}
\end{equation}
Since $\sigma_G$ has been found above (\ref{eq_sigmaG}), all that
remains is to calculate the expression:
\begin{eqnarray}
v_G&-&v_0 = 2\int_0^\infty \frac{\left(\cos(Gr)-1\right)}{(r+a)^\lambda}\dr \nonumber\\
&=& 2G^{\lambda-1} \Gamma(1-\lambda)
\sin\left(\frac{\pi\lambda}{2}\right) \quad\lambda\in[1,2)~,
\label{eq_vg2}
\end{eqnarray}
where the second line, again obtained by taking the limit
$a\rightarrow 0$ and true only for $1 \le \lambda < 2$, is precisely
the same as (\ref{eq_vg}) for $\lambda < 1$.  Since $\sigma_G$
agrees with $\sigma'_G$ except at $G=0$, and $G=0$ is excluded in
the sum of (\ref{eq_Edensity11}), the results obtained for $\lambda
< 1$  and $1 \le \lambda < 2$ separately may be combined, expanding
the region of validity of (\ref{eq_optimalCase1}) to $0 \le \lambda
< 2$.

\begin{figure}[t]
\begin{center}
\includegraphics[width=2.8in]{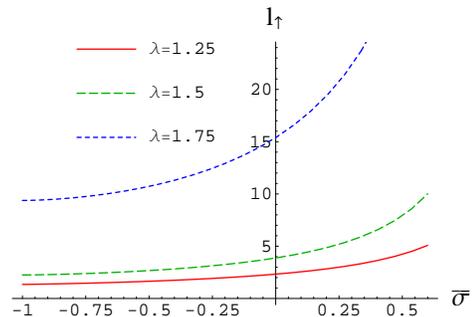}
\caption{Size of the spin up domain in the ground state of
(\ref{eq_Ham2}) as a function of average spin $\sbr$. The vertical axis is in units of $\left(\frac{2-\lambda}{\lambda}\frac{J}{Q}\right)^\frac{1}{2-\lambda}$.}
\label{figLoptVariousLambda}
\end{center}
\end{figure}

\begin{figure}[b]
\begin{center}
\includegraphics[width=2.8in]{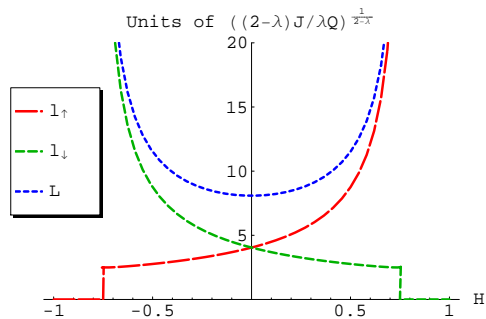}
\caption{Size of the spin up domain $l_\uparrow$, spin down domain $l_\downarrow$,
and their sum  $L$ (period of the configuration) that minimizes the
energy of classical Hamiltonian (\ref{eq_Ham2}) with $\lambda=1.5$.
We obtain these quantities as a function of applied
field $H$ by performing a Legendre transformation.\label{figAllLs}}
\end{center}
\end{figure}

In figure \ref{figLoptVariousLambda}, the size of the up spin domain
($l_\uparrow$) is plotted at several values of $\lambda \in (1,2)$
as a function of $\sbr$.  As $\lambda$ increases, the frustrating
interaction becomes shorter ranged, and the size of the domains
becomes larger as expected.  One must be careful, however, in
interpreting figure \ref{figLoptVariousLambda} since the unit itself
depends on $\lambda$.  This exposes a difficulty of working with the
continuum model with a variable exponent $\lambda$, as the only
length scale is dependent both on $\lambda$ and $J/Q$.  Figure
\ref{figAllLs} shows for fixed $\lambda=3/2$ the size of both spin
domains, as well as their sum $L$, as a function of the applied
field $\rmH$ (obtained via Legendre transform).



\subsection{Case III. \,$\lambda>2$}
When $\lambda > 2$, to keep the energy density finite, the
ultraviolet cutoff $a$ cannot be set to zero, and $a$ must be
carried through the calculation. Although the analysis leading to
(\ref{eq_Edensity11}) and (\ref{eq_vg2}) can be extended to $\lambda
> 2$, the sum in (\ref{eq_C}) would diverge, and it is preferable to
analyze the system entirely in real space, starting from
(\ref{eq_Edensity10}). Only regions where $\sigma(r) \ne \sigma(r')$
(i.e. $\sigma(r)\sigma(r')=-1$) yield nonzero contributions in the
second term, and after some algebra, for $L \gg a$ the energy density can be
written as:
\begin{equation}
\epsilon(\sbr,L) = \frac{4J}{L} +
\frac{4Q}{(\lambda-2)(\lambda-1)L}\left[
\frac{C'(\sbr,\lambda)+\alpha^{2-\lambda}}{L^{\lambda-2}} -
\frac{1}{a^{\lambda-2}} \right]
\end{equation}
where we have defined the following sum, convergent for $\lambda > 1$:
\begin{equation}
C'(\sbr,\lambda) = \sum_{n=1}^\infty \left[ (n+\alpha)^{2-\lambda} -
2n^{2-\lambda} + (n-\alpha)^{2-\lambda} \right]~, \label{eq_defC1}
\end{equation}
and $\alpha=(1+\sbr)/2$.  For larger $a/L$ there are corrections to
$C'$ of order $a/L$.  We set the derivative of the energy per site
with respect to the period $L$ to zero and thus find the ground
state $L$:
\begin{eqnarray}
\frac{\mathrm{d}\epsilon}{\mathrm{d}L} & = & -\frac{4J}{L^2} -
\frac{4Q}{(\lambda-2)L^\lambda}\left(C'(\sbr,\lambda)+\alpha^{2-\lambda}\right)
\nonumber \\ & & +\,
\frac{4Q}{(\lambda-1)(\lambda-2) L^2}a^{2-\lambda} \nonumber \\
0 & = & -J + \frac{Q}{\lambda-2}\left[ \frac{a^{2-\lambda}}{\lambda-1}-L^{2-\lambda}\left(C'(\sbr,\lambda)+\alpha^{2-\lambda}\right) \right] \nonumber\\
\frac{J}{Q} & = & \frac{1}{\lambda-2}\left[
\frac{a^{2-\lambda}}{\lambda-1} -
\frac{\alpha^{2-\lambda}C'(\sbr,\lambda)}{L^{\lambda-2}} \right]
\,.\label{eq_lambdaG2}
\end{eqnarray}
In the present case of $\lambda > 2$, $C'(\sbr,\lambda) > 0$, so the
last line above will have a solution with finite L only below a
critical value of $J/Q$:
\begin{eqnarray}
\left(\frac{J}{Q}\right)_c = 
\frac{a^{2-\lambda}}{(\lambda-2)(\lambda-1)} \,. \label{eq_maxJQ}
\end{eqnarray}
For larger values of $J/Q$ the energy is minimized by infinite $L$
so the ground state has macroscopic phase separation, since the
energy of a domain wall is then always positive.

\begin{figure}[H]
\begin{center}
\includegraphics[width=2.8in]{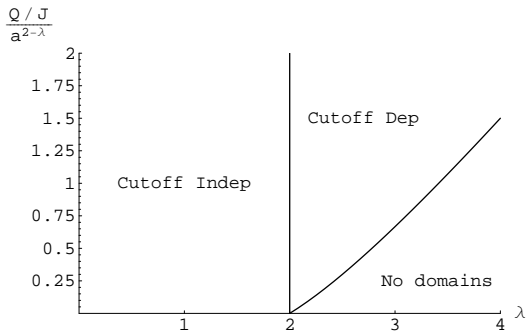}
\caption{Zero temperature phase diagram of Hamiltonian
(\ref{eq_Ham2}), showing where finite domains are present in the
ground state, and whether the domain sizes are dependent on the UV
cutoff $a$ in the limit $a \rightarrow 0$.\label{figPhaseDiag}}
\end{center}
\end{figure}

Thus, for $\lambda > 2$, this model with a given $J/Q$ may or may not
show finite domains in its ground state, depending on the cutoff
$a$. In this regime, finite domains form in the ground state only
for $J/Q < (J/Q)_c$, as specified by (\ref{eq_maxJQ}). Above this
value the ferromagnetic interaction is too strong relative to the
antiferromagnetic interaction for domains to form. Figure
\ref{figPhaseDiag} shows the regions of phase space where
domains exist, and where the domain size is cutoff independent in
the limit $a \rightarrow 0$.

\subsection{Case IV. \,$\lambda=2$}
In the special case $\lambda = 2$, an analysis similar to the above $\lambda > 2$ case can be done, which results in an energy density (valid for $L \gg a$)
\begin{eqnarray}
\epsilon(\sbr,L) &=& \frac{4J}{L} +
\frac{4Q\beta^2}{L} \left[\ln\left(\frac{a}{\alpha L}\right)-\frac{\alpha L}{a} \right]\nonumber\\
 & & + \frac{4Q\alpha^2}{L} \left[\ln\left(\frac{a}{\beta L}\right)-\frac{\beta L}{a} \right] \nonumber\\
 & & + \frac{8Q\alpha\beta}{L}\ln\left( \frac{a}{\alpha\beta L}\right) + \frac{F(\sbr)}{L}
\end{eqnarray}
where $\beta = (1-\sbr)/2$ and $F(\sbr)$ is a function independent of $L$.  By solving
$\frac{\mathrm{d}\epsilon}{\mathrm{d}L}=0$ we find that the ground
state period $L_0$ is exponentially
dependent on $J/Q$ with a prefactor proportional to $a$:
\begin{equation}
L_0 = a\exp\left[ \frac{J}{Q}+F'(\sbr) \right]
\end{equation}
where $F'(\sbr)$ is another function independent of $L$ and of order unity for $\sbr \in [-1,+1]$. This form can be argued from dimensional analysis, since $J/Q$ is
dimensionless at $\lambda=2$, and the only length scale in the
problem is $a$.  The boundary case $\lambda=2$ separates the regime
where $L$ approaches a finite limit given by (\ref{eq_optimalCase1})
as $a \rightarrow 0$ for $\lambda < 2$, from the the
cutoff-dependent regime for $\lambda > 2$.

\section{Concluding Remarks}
In summary, we have studied the generalized Coulomb-frustrated Ising
model in a one dimensional continuum for different exponents
$\lambda$ of the long range interaction ($\lambda = 1$ for the
Coulomb case). We have derived an analytic solution for the ground
state domain configuration of this model under the assumption that
the ground state configuration has a simple periodic structure. This
assumption has been confirmed by Monte Carlo simulations (see
Appendix \ref{appMonteCarlo1D}).  Such simulations can be done in
any dimension $d$, and mark an avenue for future work; more
complicated domain patterns presumably do occur for $d>1$.

We find that for $0 < \lambda < 2$, as the magnetic field $\rmH$ is
increased from $-\infty$, the ground state is ferromagnetic until a
critical field $-\rmH_c$ is reached, at which point isolated domains
of flipped spins of finite length are formed in an otherwise
polarized background. For $-\rmH_c \le \rmH \le \rmH_c$, periodic
configurations with $l_\uparrow$ up spins followed by $l_\downarrow$
down spins become the ground state, and the system is said to be in
a ``microemulsion'' phase.  In higher dimensions much numerical work
has been done in the absence of a magnetic
field(see\,\cite{LowEmery,ViotTarjus,MuMa}), and applications have
been made to magnetic thin films\cite{DeBellRMP} in 2D, as well as
the metal insulator transition in 2D and 3D\cite{Ortix2006}.
Analytical expressions analogous to our results can be derived
assuming the ground state is simply periodic along only one
direction\cite{NgVanderbilt}.  But more generally one is presumably
forced to resort to approximations and numerics.  The numerical
Monte Carlo work, however, can be readily extended to higher
dimensions, and can be used to investigate domain formation and
behavior as a magnetic field is varied (\emph{i.e.} in the
non-charge-neutral case).

Our solutions describe how $l_\uparrow$ increases and $l_\downarrow$
decreases with increasing $\rmH$.  At zero field $l_\uparrow =
l_\downarrow$, and at $\rmH=\pm\rmH_c$ we find that $l_\uparrow$ or
$l_\downarrow$ diverges to infinity, respectively, while the length
of the ``minority'' domain remains nonzero and finite.  For $\lambda
\ge 2$, whether or not the microemulsion phases appear in the
transition between fully polarized up and down states depends on the
dimensionless quantity $(Qa^{\lambda-2})/J$. In the case that
microemulsion phases do occur in this region, their properties
depend on the ultraviolet cutoff $a$.  
Our results provide more explicit examples of models with
frustrating and sufficiently long range interactions that have
ordered ``microemulsion'' phases instead of macroscopic phase
separation. However, whether such a result is obtained in an
inherently \emph{quantum} system such as the 2D electron gas
requires further investigation. In this context, it is worth noting
that claims exist in the literature in favor of the classical
scenario, both for the electron gas\cite{Spivak2004} and for the
highly disordered Anderson model with long range Coulomb
interactions\cite{BhattRamakrishnan}.


\appendix

\section{Monte Carlo analysis \label{appMonteCarlo1D}}
The approximation central to this paper is that the ground state of $\mathcal{H}_{\sbr}$ (see eq. \ref{eq_Ham2}) has a simple periodic structure.  While this has been proved for $\rmH=0$ (\emph{i.e.} $\sbr=0$), no such result exists for nonzero magnetization.  Thus, to justify the approximation, we have performed Monte Carlo simulations on a discretized version of eq. \ref{eq_Ham2}:
\begin{equation}
\mathcal{H}_{\sbr} = -J\sum_i \sigma_i \sigma_{i+1} + Q\sum_{ij}
\frac{\sigma_i\sigma_j}{|i-j|^\lambda}~.
\end{equation}
If we introduce a lattice spacing parameter $a$, then in the limit $J/Q \rightarrow 0$ and $a \rightarrow 0$ such that $\frac{Q a^{\lambda-2}}{J}=C$ for constant $C$, the above Hamiltonian is equivalent to the continuum formulation (eq. \ref{eq_Ham2}) with $J/Q=C$.  We solve for the ground state the discretized model using a Monte Carlo algorithm with simulated annealing.  Updates are determined by a Wolff cluster method\cite{Grousson2001} which preserves the overall magnetization, and thus $\sbr$ is fixed during a simulation run (similar to the analytical calculation).  Enough runs are averaged over so that the variation in the sizes of the resulting domains is negligible.  Additionally, the system configuration converged upon by each Monte Carlo run is compared to the exactly periodic configuration and it is seen that the energy density of the exactly periodic configuration is always equal to or lower than the energy density of the Monte Carlo result.  In summary, we find that at any $\sbr \in [-1,1]$, and for all values of $J/Q$ the ground state consists of uniform spin up domains of length $l_\uparrow$ interleaved with uniform spin down domains of length $l_\downarrow$ (but $l_\uparrow \ne l_\downarrow$).  Detailed results will be provided elsewhere\cite{NielsenMonteCarlo}.


\end{document}